# Mapping of periodically poled crystals via spontaneous parametric down-conversion

G.Kh.Kitaeva[1*], V.V.Tishkova[1], I.I.Naumova[1], A.N.Penin[1],

C.H.Kang[2], S.H.Tang[2]

[1]Department of Physics, M.V.Lomonosov Moscow State University, 119899 Moscow, Russia

[2]Department of Physics, National University of Singapore, Singapore 117542

**Abstract.** A new method for characterization of periodically poled crystals is developed based on spontaneous parametric down-conversion. The method is demonstrated on crystals of Y:LiNbO$_3$, Mg:Y:LiNbO$_3$ with non-uniform periodically poled structures, obtained directly under Czochralski growth procedure and designed for application of 1.064 μm-pumped OPO in the mid infrared range. Infrared dispersion of refractive index, effective working periods and wavelengths of OPO were determined by special treatment of frequency-angular spectra of spontaneous parametric down-conversion in the visible range. Two-dimensional mapping via spontaneous parametric down-conversion is proposed for characterizing spatial distribution of bulk quasi-phase matching efficiency across the input window of a periodically poled sample.

## 1. Introduction

During the last two decades, the use of periodically poled (PP) crystals has greatly extended the area of new nonlinear optical ideas and devices [1,2]. At the same time, various methods for fabricating PP elements are widely developed [3] and the problem of characterization for such PP structures becomes essential. Usually, the PP structures have the desired periods and directions with respect to crystal axes, but the degree of their periodicity is limited, and domain walls can be non-planar; other defects occur depending on the poling technique. Evidently, these effects decrease non-linear efficiencies of PP elements, and the methods for estimating final efficiencies are crucial.

---

[*] Fax: +7-095/939-1104, E-mail:kit@qopt.phys.msu.su

The most prevalent method of characterizing PP structures is imaging of chemically etched crystal surfaces [4], based on unequal etching rates for domains of anti-parallel polarization. As other surface-based methods, such as electro-optic imaging microscopy [5], scanning electron microscopy [6], near-field scanning optical microscopy [7], scanning atomic force microscopy [4], x-ray diffraction imaging [8], optical microscopy of etched samples enable us to measure directly sizes of domains along a selected plane. Nevertheless, only a limited area of the structure is studied, the inner parts are inaccessible without destruction of the sample. One needs to make a statistical treatment of the data on domain widths, obtained from the whole surface of the structure, to make conclusion about its quality for a nonlinear optical application. On the other hand, the methods based on non-linear optical effects such as imaging via second-harmonic generation [9-11], domain mapping via terahertz wave form analysis [12], and the method based on spontaneous parametric down-conversion (SPDC) [13-15] are free of these disadvantages. Usually, an appropriate non-linear effect is taken as an initial probing process to test inner parts of a PP structure and to predict its general characteristics in other non-linear applications. Among these methods, SPDC provides information for the widest spectral range, and can be organized in different quasi-phase matching regimes for the largest variety of PP crystals. This is due to the fact that a large reservoir of zero vacuum fluctuations acts during SPDC process as probe idler waves of all possible directions and in the whole spectral range below a pump frequency [16].

In this paper, we further develop the SPDC-based characterization technique, for the first time propose the SPDC-based mapping of PP samples, and demonstrate the ability of this method by the example of PP crystals of doped lithium niobate (PPLN), Y:LiNbO$_3$ and Mg:Y:LiNbO$_3$, designed for applications in OPO schemes for generation of mid infrared (MIR) radiation near 4 µm. The crystals contained 1-dimensional domain grating produced directly under Czochralski growth procedure [17]. This enabled PPLN samples of larger transverse sizes, with domain walls of higher threshold upon possible thermal and radiation influence [18-19] in comparison with PPLN crystal chips obtained under post-growth techniques [1,4,21,22]. Nevertheless, degree of periodicity is not so high everywhere and can be different in different parts of the grown bole. The initial characterization and selection of the best part is much desired so as to derive maximum efficiency in further applications. Apart from the applications, where some special types of aperiodicity are necessary (for example, Fibonacci- or Cantor-type nonlinear gratings [23-25], stochastic disordered structures [2,25-27]), a higher regularity is required. Thus, it is appropriate to select the regular parts by analyzing the visible nonlinear signal, and then to recommend this part for applications in other spectral ranges. For example, tuning curves for quasi-synchronous *eee*-type MIR OPO are known to be very sensitive to dispersion of refractive index and to period of nonlinear grating [28]. This fact further strengthens the need for greater accuracy in characterization.

The paper is organized as follows. Sec. 2 contains requirements for the samples from the viewpoint of MIR-range generation, details of the samples fabrication procedure, and results of their primary study by chemical etching. Further these results are compared with the data of SPDC–based characterization, which consist of three steps. Obtained as the first step of characterization, overview SPDC spectra are presented in Sec. 3. In Sec. 4, we discuss how such spectra are formed in crystals with periodic and non-periodic spatial distribution of second-order optical susceptibility and present results of their treatment at the second characterization step. SPDC-based mapping, made finally at the third step, is considered in Sec. 5. Results are summarized in Sec. 6.

## 2. Samples

The samples were cut from 2 boules of bulk PPLN, with chemical formula Y:$LiNbO_3$ and Mg:Y:$LiNbO_3$. The boules were grown using Czochralski technique along X-axis from close to congruently melting composition Li/Nb = 0.942 with 1 wt % $Y_2O_3$ and, in case of Mg:Y:$LiNbO_3$, with 2 mol% MgO. Doping by Y was used to form PP domain structure on the basis of rotation-induced growth striations in the bulk of $LiNbO_3$ single crystals. Codoping by Mg, not only reduces photorefractive damage, provides additional fixing of domain walls at the growth striations. As it is established for this domain fabrication technique, period of a nonlinear grating is determined by the ratio between pulling and rotation rates under the growth procedure [17,29]. In our case pulling and rotation rates were 10 mm/h and 6 rpm, correspondingly, the expected value of PP structure period was 28 μm. The samples of 17-18 mm long were cut normally to the crystal X-axis from central parts of the boules. Parallel to domain walls, the input and output surfaces of the samples were polished. These surfaces were of 6-10 mm along Z-axis, and of 2-3 mm along Y-axis.

The primary view of the ferroelectric domain structure was studied in all samples by selective chemical etching. The crystal surfaces normal to Z-axis were polished and subsequently etched in 1:2 v/v mixture of HF and $HNO_3$ acid in Pt-crucible for 2-3 min at boiling temperature. We observed the profile of the etched surfaces and made optical images of them using a digital camera and an optical microscope Nikon MM-40. Examples of optical micrographs are shown in Fig.1a,b. Analysis of surface images for Mg:Y:$LiNbO_3$ crystal shows that the structure is rather regular in the middle parts of the sample surface, but some irregularities can be noticed in parts located near peripheral areas of the bole. As it was expected, the degree of periodicity in Y:$LiNbO_3$ crystal samples was not uniform in all parts of their surfaces (Fig.

1b). Usually, the etching technique enables one to measure periods of the PP structure at Z surfaces, but may be inconsistent in estimating the thicknesses of positive and negative domains because of too deep etching, which masks a real ratio of the thicknesses. Relative thicknesses of positive and negative domains in our samples were studied by electric force microscopy(Fig.1c). The distribution of electric field at Z-surfaces indicated that adjacent domain layers were of the same width and a duty cycle of 50% was characteristic for the structures on average.

The typical periods of nonlinear gratings were found to be within 32-33 μm for Y:Mg:LiNbO$_3$ crystal samples and within 30-34 μm for Y:LiNbO$_3$ samples. The increasing of the period size as compared with calculated period (28 μm) was connected with decreasing of the melt level in the crystal growth process. Nevertheless, the values beyond these intervals were also observed. It became obvious that laborious statistical treatments of images via the whole sample lengths would be necessary for formulating any reliable conclusions about the most effective average period for each sample. At the same time, necessary requirements for parameters of nonlinear gratings are usually quite strict. Using available data [30-34] on refractive index dispersion for pure congruent and Mg-doped crystals, we determined wavelengths of MIR idler radiation near 4 μm under *eee*-type parametric down-coversion of pump radiation at 1.064 μm in PPLN and Mg:PPLN crystals with nonlinear gratings of different periods (Fig.2). It was shown that

1) when period of a non-linear grating changes very slightly - on 0.1 μm, the wavelength of idler radiation in MIR is shifted appreciably - to about 0.035 μm;
2) temperature tuning in the range 20-200 $^0$C changes the idler wavelength up to about 0.055 μm (increase of the temperature leads to decrease of the MIR wavelength);
3) decrease of Mg-content leads to decrease of an idler wavelength, which can be compensated by proper decrease of a grating period.

In order to clarify how much the samples satisfy the conditions necessary for non-linear optical applications, the SPDC-based characterization was utilized.

## 3. Observation of frequency-angular SPDC spectra

Fig.3 illustrates the principal scheme of SPDC-spectrograph - the experimental set-up for registration of two-dimensional frequency-angular spectra of SPDC in nonlinear crystals [35]. The linearly polarized pump beam of laser source is incident on a crystal. The scattered radiation passes through an optical registration system, which includes an analyzer, lenses, and a spectrograph, to be finally recorded by a scanning detector or by a photographic film mounted at the spectrograph output window. SPDC process in any medium with second-order susceptibility can be treated as a spontaneous decay of pump photons of frequency $\omega_p$ into signal and idler photons of frequencies $\omega_s$ and $\omega_i$: $\omega_p = \omega_s + \omega_i$ [16]. In spectroscopic schemes, $\omega_p$ and $\omega_s$ are in a spectral region of crystal transparency, usually, in a visible range; idler frequencies $\omega_i < \omega_s$ and can occupy the IR transparency region, as well as the range of phonon absorption. When $\omega_i$ hits the spectral range of optical phonons, SPDC is transformed into near-forward Raman scattering by polaritons, which is also observable via SPDC-spectrograph [35-38]. In SPDC-spectrograph the signal radiation is sorted according to its frequency and according to its angle of direction. Usually, a frequency sweep is made along the axis, which is oriented normally to a spectrograph entrance slit. An additional angle sweep is made along the transverse axis, which is parallel to the slit. The angle sweep is arranged by a lens system, which focuses the signal radiation on the slit. Displacement of the focus position from the center of the slit is proportional to $tg\theta_s$, where $\theta_s$ is the angle of the signal direction with respect to the optical axis of the system (Fig.3).

Fig.4 and Fig.5a show the examples of SPDC overview spectra obtained for our samples in *ooe*-geometry: pump was directed along crystallographic X-axis and extraordinarily polarized along Z-axis, the idler and signal waves were directed in XZ plane and ordinarily polarized along Y-axis. Ar-laser radiation at a wavelength of 488 nm was used as a pump beam. The horizontal spectral sweep is given in terms of signal wavelengths and represents the parts of the whole spectra which correspond to SPDC in transparency region (in other words, to near-forward Raman scattering at the upper polariton branch). Here the idler frequencies are more than the largest longitudinal phonon frequency: $\nu_i = \omega_i/2\pi c > 880$ cm$^{-1}$. Formally, the region of crystal transparency begins from this side, phonon-induced absorption decreases when the idler frequency is further increased. In this region the SPDC spectra consist of curves, which display actually the quasi-synchronous tuning curves valid for any types of parametric three-wave mixing processes (parametric generation, amplification, frequency conversion, etc.) in the corresponding crystal, type of polarization, and frequency conditions.

## 4. Characterization of crystals via SPDC spectra

As it was shown theoretically [39], distribution of signal intensity across SPDC spectrum $P(\omega_s, \theta_s)$ is described by general relation, which presents the signal intensity as a function of the wave mismatch in any three-wave mixing process at its initial low-amplification stage. If absorption is negligible, it takes a simple form,

$$P(\omega_s, \theta_s) = C_0 \left| \sum_{m=-\infty}^{\infty} \chi_m^{(2)} \text{sinc}(\Delta_m/2) \right|^2. \quad (1)$$

where $C_0$ is a slowly varying coefficient, $\chi_m^{(2)}$ are amplitudes of spatial Fourier harmonics,

$$\chi_m^{(2)} = \frac{1}{d} \int_{-d/2}^{d/2} \chi^{(2)}(x) e^{-imqx} dx, \quad (2)$$

describing spatial variation of the second-order susceptibility $\chi^{(2)}(x)$ in terms of Fourier series:

$$\chi^{(2)}(x) = \sum_{m=-\infty}^{\infty} \chi_m e^{imqx}. \quad (3)$$

Here X-axis is directed normally to domain layers. $\Delta_m \equiv \Delta_m(\omega_s, \theta_s)$ are dimensionless quasi-phase mismatches of different orders m, they are defined as X-components of vector quasi-phase mismatches. In the case of SPDC,

$$\Delta_m(\omega_s, \theta_s) = \left[ (\mathbf{k}_p - \mathbf{k}_i - \mathbf{k}_s) - m\mathbf{q} \right]_x L. \quad (4)$$

$\mathbf{k}_p$, $\mathbf{k}_i$, and $\mathbf{k}_s$ are wavevectors of pump, idler and signal waves in a crystal, correspondingly; $\mathbf{q}$ is the vector directed along X and characterizing an inverse nonlinear superlattice; $|\mathbf{q}| = 2\pi/d$, where d is the period of a spatial variation of the second-order susceptibility $\chi^{(2)}(x)$. For a non-periodic spatial $\chi^{(2)}(x)$ variation, d equals to a whole crystal thickness L along the normal to domain layers X.

Relation (4) enables one to determine quasi-phase mismatches, corresponding to each point $(\omega_s, \theta_s)$ on a SPDC spectrum, but the data on dispersion of refractive indexes is necessary to calculate the wavevectors. For the first approximation, we used Sellmeier formulas, obtained in [31] for pure congruent LiNbO$_3$ while calculating the spectra of Y:LiNbO$_3$. For Mg:Y:LiNbO$_3$ we used our previous data on refractive index dispersion in Mg:Nd:LiNbO$_3$ [33]. Mg-doping concentrations were the same in both cases, but rare-earth elements were different. Since they enter crystals at very low concentrations, it seems justified to expect their influence on refractive indexes to be small. Following these assumptions, we have determined, which closed tuning curve is the result of synchronous interaction, characterized by m=0

($\Delta_0 = 0$). Next, we found that the whole group of low-intensity closed curves can be attributed to quasi-synchronous interaction of order m=-2 ($\Delta_{-2} = 0$), the two other groups, containing rather intense curves – to m=-1 or m=+1 ($\Delta_{\pm 1} = 0$). This attribution enabled us to obtain precise data on the refractive index dispersion for the Mg-doped crystal in the IR range. Results for ordinary refractive index dispersion under doping level of 2 mol % can be fitted by formula

$$n_o^2(\lambda) = A - \frac{B\lambda^2}{C - \lambda^2} + \frac{D}{\lambda^2 + E}, \qquad (4)$$

for MIR wavelengths λ between 1.5 μm and 10.5 μm within the mean accuracy of ±0.001; A=4.8506, B=10.824, C=434.57, D=0.3355, and E=1.945. It was shown, that Y-doping in our crystals, as well as Nd-doping in [33], do not affect refractive index dispersion within the given accuracy of measurement.

Different curves in each group at SPDC spectrum correspond to the same order of quasi-synchronous interaction in nonlinear gratings with slightly different periods *d*. All these values of $d_1, d_2....$ were within the intervals obtained by measuring the etched surfaces. Such type of SPDC spectra is representative of an irregular distribution of domain thickness, when the structure period in (1) is taken equal to the length of the whole crystal L. Special treatment shows [14,15] that one can measure the relation between modules of different-order harmonics $\left|\chi_m^{(2)}\right|/\left|\chi_n^{(2)}\right|$ as $\left|\chi_m^{(2)}\right|/\left|\chi_n^{(2)}\right| = \sqrt{I_m/I_n}$, where $I_j (j = m, n)$ are intensities of the different-order curves corresponding to conditions $\Delta_j(\omega_s, \theta_s) = 0$, i.e. $(\mathbf{k}_p - \mathbf{k}_i - \mathbf{k}_s)_x L = 2\pi j$. By making an inverse Fourier treatment (2), one can determine the whole distribution $\chi^{(2)}(x)$ in a sample along the pump beam [14]. Nevertheless, this inverse treatment procedure is not quite necessary when the sample is characterized for further applications. Indeed, in other three-wave interactions, in the initial low-amplification stage, the intensity dependence on the phase mismatch will be of the same character, assigned by exp. (1). Only the magnitude (4) of the mismatches as functions of frequencies and/or angles will be modified in accordance with new dispersion characteristics of the crystal in another frequency range. We can thus regard a crystal sample as a number of built-in nonlinear gratings, characterized by different working periods with their individual non-linear efficiencies. The characteristic efficiency is described by a square of corresponding $\chi_m^{(2)}$ here, which determines signal intensity $I_m$ in a quasi-phase matching condition $\Delta_m = 0$.

Distribution of the efficiencies along the equidistant spectrum of inverse vectors $2\pi m/L$, $2\pi(m+1)/L,...$ can be dissimilar in different parts of the sample. As a result, the generation at corresponding idle wavlengths can be of different intensity in further application. Fig.5b shows the SPDC intensity as a function of the signal wavelength, measured for one of our samples in collinear direction ($\theta_s$=0). The corresponding values of dimensionless mismatch $\Delta(\omega_s,\theta_s) = \left[(\mathbf{k}_p - \mathbf{k}_i - \mathbf{k}_s)\right]_x L/2\pi$ are indicated at the top axis. The scale of signal wavelengths under SPDC can be re-counted also in general terms of working periods $d_{eff} = L/\Delta(\omega_s,\theta_s)$, and, if necessary, in terms of output wavelengths under a special application. The intensity scale, taken in relative units, describes the efficiency of the corresponding working periods. As for example, the corresponding values of MIR idler wavelengths for 1.064 μm-pumped OPO are indicated at additional horizontal axis in Fig.5b.

## 5. SPDC mapping

Each SPDC spectrum can be used to characterize that part of a sample volume, which has been a source of a signal measured under SPDC, and is regarded as a source in other applications. In case of our SPDC spectra, it had transverse sizes of a pump beam and was as long as a whole sample. A set of overview spectra, obtained for different parts of each crystal sample, showed, that distribution of efficiencies among various working periods is different. To characterize the whole volume of the samples, we organized the special SPDC-based mapping procedure.

The polarized Ar-laser pumping beam of 1 W was focused on a sample with a spot size of about 50 μm along the whole sample length. The sample was mounted on YZ-motorized stage for subsequent scanning of the pumping spot across the sample. The forward SPDC signal at $\theta_s$=0 was selected using diaphragms, analyzer, and filter, and detected by a Jobin Yvon T64000 spectrometer for each position of the spot with a dwelling time of 60s. The numerous curves such as examples in Fig.5b were measured. After subtracting the noise background, the intensity of the peak at 603nm was measured as a function of crystal position. This peak corresponds to the $\chi_m^{(2)}$ Fourier harmonic, which determines intensity of idler wave at 4.25 μm in OPO application under pumping at 1.064 μm. The scanning was made along crystallographic Y and Z axes with steps of 0.1 mm. Finally, it enabled us to build the map of efficiency distribution across the sample surface. The map for the mostly non-uniform sample of Y:LiNbO$_3$ is presented in Fig.6. The brighter areas on the map correspond to more effective parts of the sample input window. SPDC map characterizes distribution of the sample efficiency across the input and output surfaces, accounting the integral effect of the overall sample length and help to choose the best place for further applications.

**6. Summary**

We proposed the optical method for mapping the quality of periodically polled crystals for application in quasi-phase matching regime. It is based on spontaneous parametric down-conversion and enables one to characterize the overall non-linear optical efficiency of inner parts of crystal samples. The method was probed on periodically poled crystals of Y:Mg:LiNbO$_3$ and Y: LiNbO$_3$ grown by Czochralski technique. It was shown how the working periods can be measured for non-regular nonlinear superlattices, and how relative distribution of their non-linear efficiencies can be determined. Although all the measurements are made in the visible range, results are suitable for applications in other spectral regions, which can be more difficult for direct access. As an illustration, the testing was made for a possible utilization of the samples in the MIR range. A two-dimensional map was recorded for the spatial distribution of efficiency of a selected working period across the input window of a sample, characterizing the integral effect of its overall length at any points of the input window.

The authors are grateful to N.F.Evlanova, K.A.Kuznetsov, and A.Solntsev for their help and fruitful discussions. This work is supported by a grant from DSTA (Grant No. POD0103451), EERSS programme, Singapore, grant for the leading scientific groups of Russia (No. 166.2003.02), and by grant 05-02-16278of Russian Foundation for Basic Research.

**Figure captions.**

Fig.1. Images of the etched Z- surfaces of Mg:Y:LiNbO$_3$ (a,c) and Y:LiNbO$_3$ (b) crystals, obtained via optical (a,b) and electric force (c) microscopes.

Fig.2. Mid infrared wavelength of 1.064 μm-pumped *eee*-type OPO based on congruent pure PPLN (solid line) or 2 mol % Mg-doped PPLN (dashed line) at room temperature, as a function of domain grating period d.

Fig.3. Scheme of SPDC-spectrograph for registration of two-dimensional frequency-angular distribution of signal radiation under spontaneous parametric down-conversion.

Fig.4. Photographic spectrum of SPDC in a periodically poled Mg:Y:LiNbO$_3$ crystal. Vertical axes: angles between signal and pump waves (outside a crystal), horizontal axes: wavelength of signal waves.

Fig.5. a: Photographic spectrum of SPDC in a periodically poled Y:LiNbO$_3$ crystal. b: Spectral distribution of SPDC signal intensity in the spectrum area, marked by a dashed line for two different regions of the crystal. Phase mismatches under SPDC, working periods of the irregular non-linear grating, and OPO wavelengths in MIR range, all corresponding to SPDC signal wavelengths, are given at additional horizontal axes.

Fig.6. Map of spatial distribution of collinear quasi-phase matched SPDC signal (at a wavelength 603nm) across the input window of Y:LiNbO$_3$ periodically poled crystal.

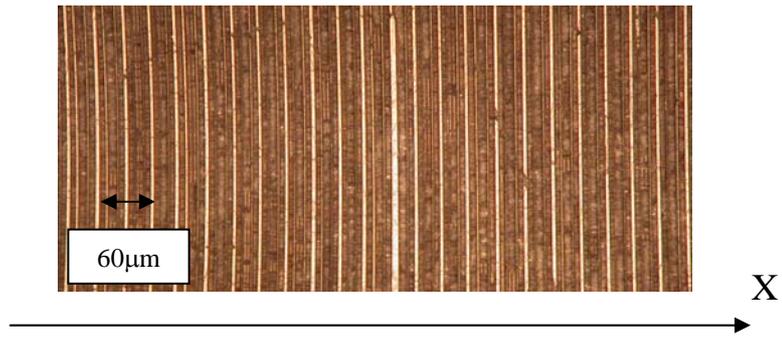

a.

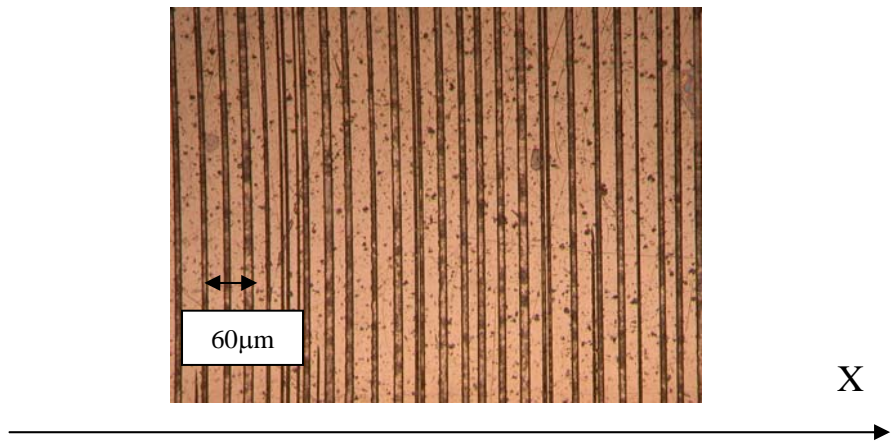

b.

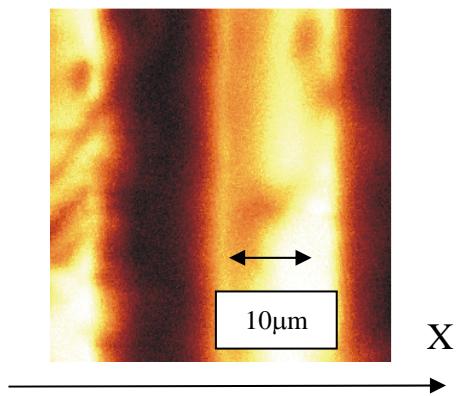

c.

Fig.1

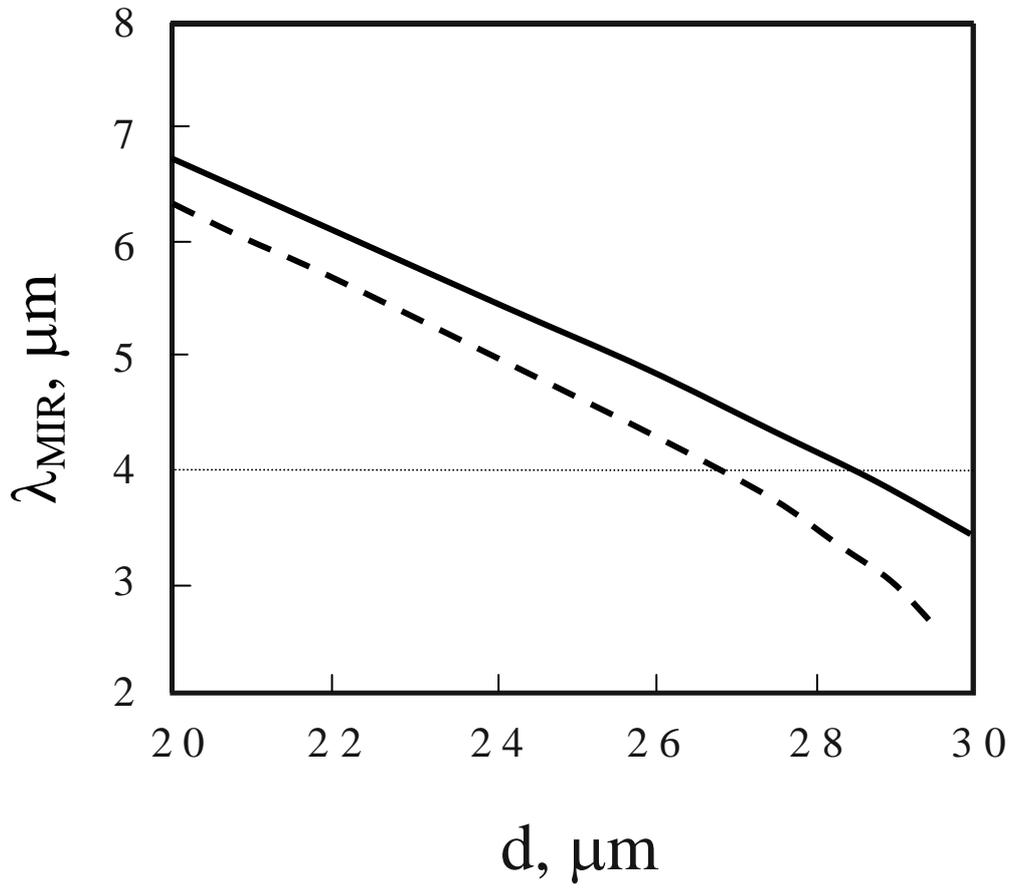

Fig.2

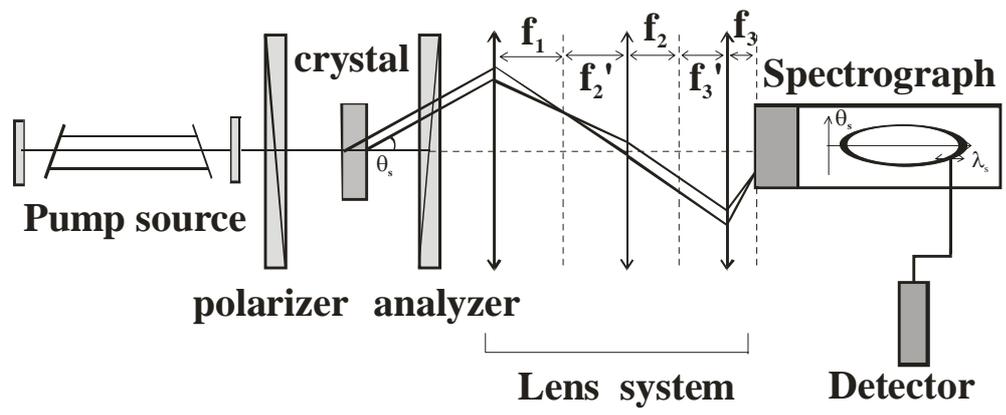

Fig.3

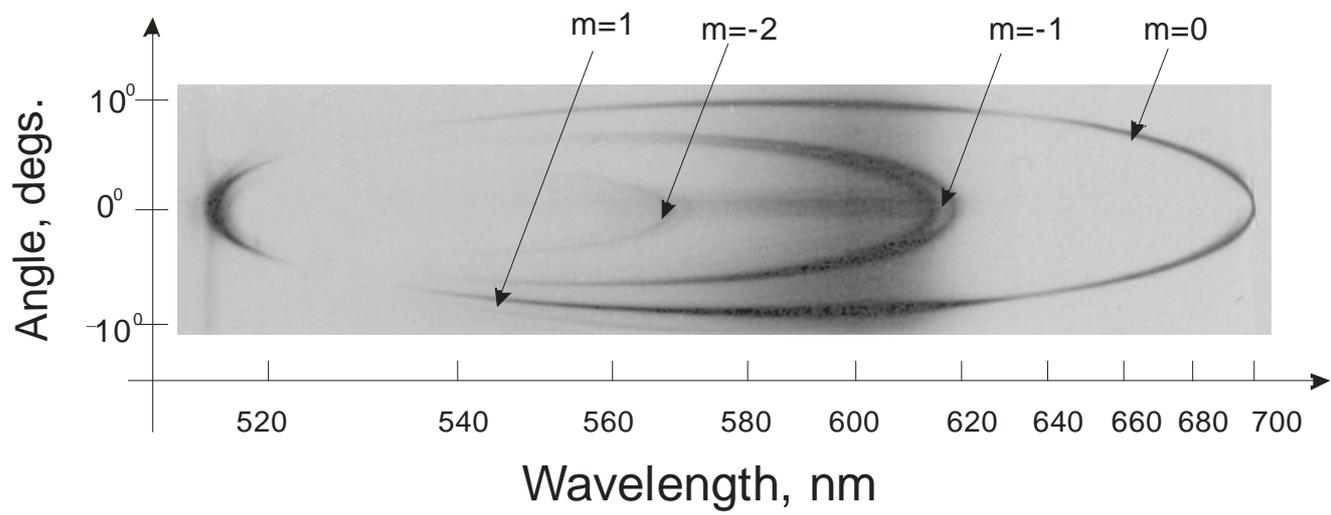

Fig.4

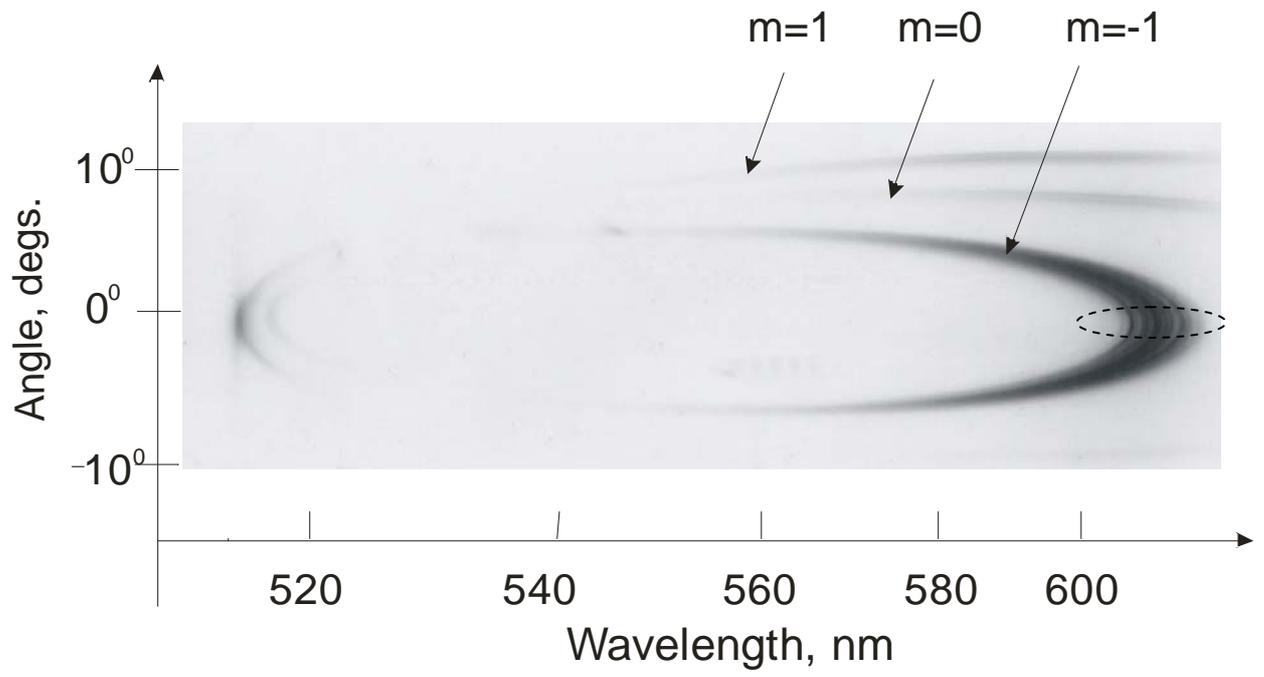

a.

Fig.5a

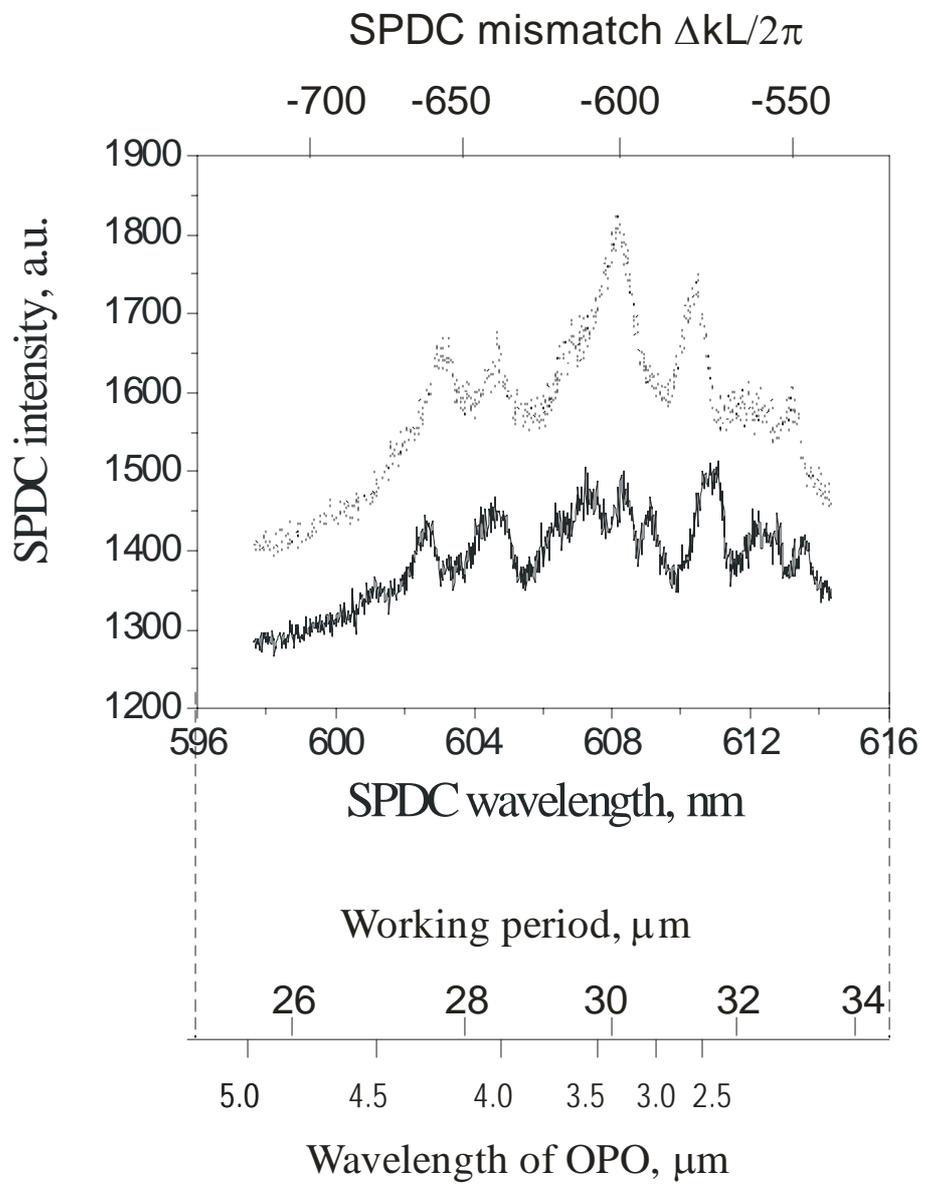

Fig.5b

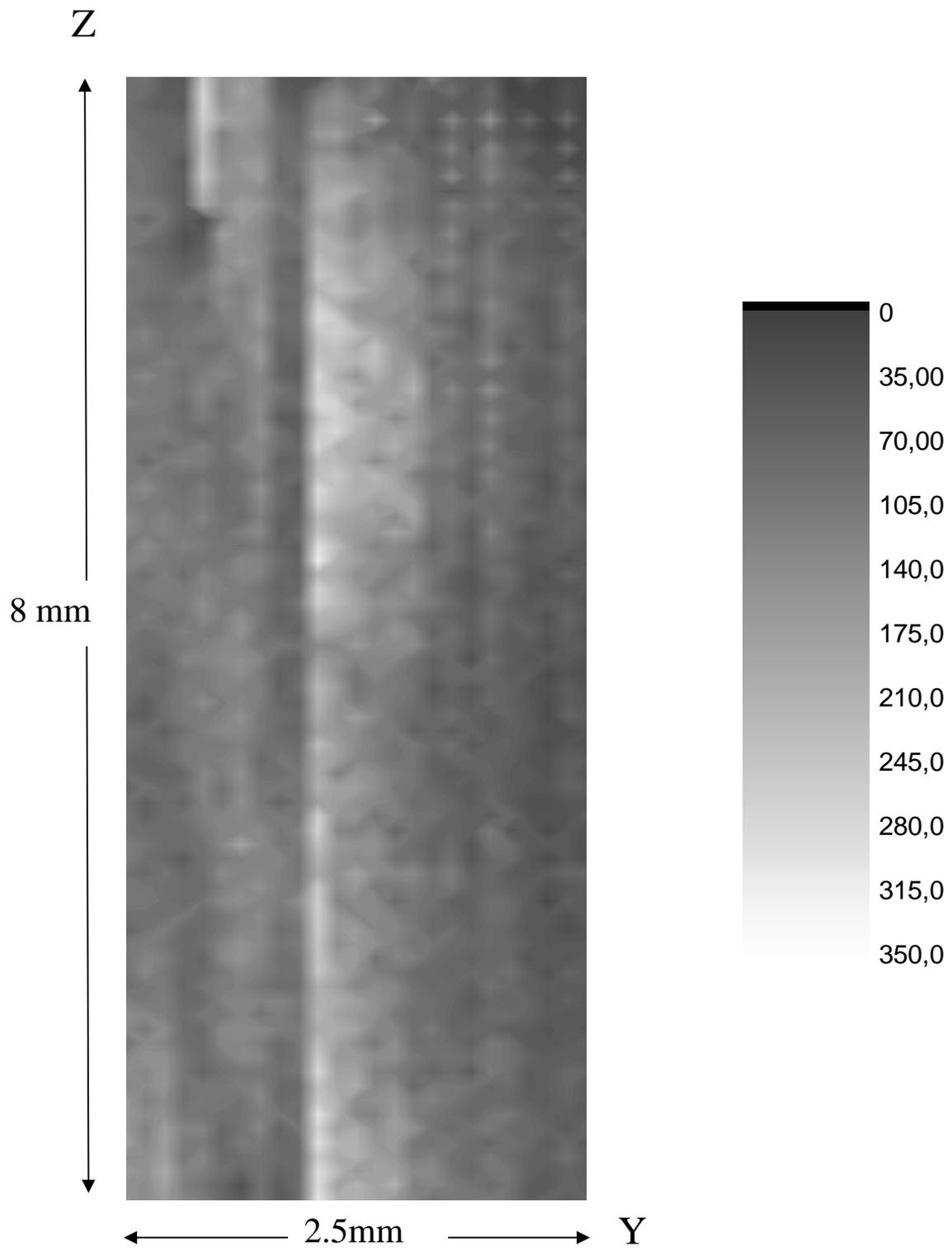

Fig.6